\documentstyle[aps,preprint,tighten]{revtex}
\begin{document}
\title{Rectangular microwave resonators with magnetic anisotropy.
Mapping onto pseudo integrable rhombus}
\author{Evgeny N.Bulgakov $^1$ and Almas F. Sadreev$^{1,2}$
\footnote{E-mail address: almsa$@$ifm.liu.se, almas$@$tnp.krasn.ru}}
\address{1) Kirensky Institute of Physics, 660036, Krasnoyarsk, Russia\\
2) Department of Physics and Measurement Technology,
Link\"{o}ping University, S-581 83 Link\"{o}ping, Sweden}
\maketitle
\begin{abstract}
The rectangular microwave resonator filed by a ferrite with uniaxial magnetic
anisotropy is considered. It is shown that this task can be reduced to
an empty rhombus resonator with vertex angle defined by external
magnetic field provided the magnetic anisotropy of the ferrite is strong.
Therefore statistics of eigen frequencies 
for TM modes is described by the Brody or semi-Poisson 
distribution with some exceptional cases.
\end{abstract}
One of the main research lines in quantum chaos is to investigate the statistics
of energy levels and eigen functions of quantum systems whose classical counterpart
is chaotic. A very popular class of systems are two-dimensional 
Euclidean billiards, which is described by the Helmholtz equation
\begin{equation}
-\nabla^2 \psi(x,y)=\lambda \psi(x,y)
\label{helm}
\end{equation}
with Dirichlet boundary conditions $\psi(x,y)=0$ for $(x,y) $ at the boundary
of billiard. It was shown that the eigen value statistics of $\lambda$ obey the
statistics of random matrix ensemble \cite{mcdonald,bohigas}. 
The distribution function for the amplitudes of the eigenfunctions 
$\psi$ is perfectly well described by a Gaussian distribution
\cite{mcdonald1,stockmann}. Correspondingly the square 
$\rho=|\psi|^2$ is describing by the well known Porter-Thomas (P-T) 
distribution \cite{stockmann}
\begin{equation}
\label{porter}
P(\rho)=\frac{1}{\sqrt{2\pi \rho}}\exp(-\rho/2).
\end{equation}

For integrable billiards, for example, rectangular one,
the eigen value statistics is describing by the Poisson distribution. 

The distribution function of $\rho$ is given by formula
\footnote{A.I.Saichev, private communication}
\begin{equation}
\label{g2d}
g(\rho)=\frac{1}{\pi^2\sqrt{\rho}}K(1-\rho), \quad 0 < \rho < 1
\end{equation}
where $K(m)$ is the elliptic integral of the first kind and the eigen state is
$\psi(x,y)=\sin(k_x x)\sin(k_y y)$. The distribution (\ref{g2d}) has a stepwise behavior
at $\rho=1$ with a value of step equaled $K(0)/\pi^2$ resulted by that 
$\rho(x,y)$ has hte same maxima.

In the two decades a subject of these distributions 
was dominated by theory and computer simulations.
About ten years ago experimentalists found effective techniques to study
billiards with the help similarly shaped electromagnetic resonators
\cite{stockmann1,sridhar,sridhar1,alt,alt1}. There are two ways to obtain 
precisely the Helmholtz equation (\ref{helm}) \cite{stockmann}. 
For the TM modes we
have $\psi(x,y)=E_z(x,y), B_z=0$ and 
$\lambda=(\omega/c)^2-(\pi n/d)^2, n=0, 1, 2,...$ where
the z-th component of electric field $E_z$ obeys the Dirichlet boundary
condition, $\omega$ is the angular frequency and $c$ is the light velocity,
$d$ is a thickness of the resonator. For the TE modes we have
$\psi(x,y)=B_z(x,y), E_z=0$ but with the Neumann boundary conditions for
$\psi(x,y)$. In what follows we consider here only the TM modes. 

   We consider more general case of the Helmgoltz equation 
\begin{equation}
\{\mu_{xx}\frac{\partial^2}{\partial{x^2}}+
\mu_{yy}\frac{\partial^2}{\partial{y^2}}+
(\mu_{xy}+\mu_{yx})\frac{\partial^2}{\partial{x}\partial{y}}+\lambda\}\psi(x,y)=0.
\label{anishelm}
\end{equation}
with the Dirichlet boundary conditions where components of tensor
$\mu_{\alpha,\beta}$ will be defined below. Because of term
with mixed partial derivatives in (\ref{anishelm}) even for 
the rectangular billiard the solution of this equation can not be presented as
$\psi(x,y)=\psi(x)\psi(y)$ and, therefore, even the rectangular billiard is becoming 
non integrable one. Below we will show that Eq.(\ref{anishelm}) for the
rectangular billiard 
can be transformed to the isotropic Helmholtz equation (\ref{helm}) but 
with the Dirichlet boundary conditions at the rhombic billiard. 
Eq.(\ref{anishelm}) describes
microwave resonator filed by anisotropic magnet. Using of
magnets in microwave resonators  was described already as a way to violate
time reversal symmetry \cite{so,stoffregen}. Here we consider a 
complete filling of resonator to have homogeneous case for which time-reversal
symmetry takes place \cite{stockmann}.

Let us write the Maxwell equations for TE modes of electromagnetic fields
in two-dimensional resonators 
\begin{eqnarray}
\label{maxwell}
\nabla\cdot{\bf B} =0,\nonumber \\
\nabla\times {\bf n}E_z = -ik{\bf B},\nonumber \\
\nabla\times{\bf H} = ik{\bf n}E_z,\nonumber\\
{\bf B}=\hat{\mu}{\bf H},
\end{eqnarray}
where ${\bf n}$ is unit vector parallel to electric field directed , ${\bf H}$ is the magnetic
field, ${\bf B}$ is the magnetic induction, $k=\omega/c$ and
$\omega$ is a eigen frequency with wave number $k$.
In what follows we imply the following magnetic 
anisotropic permeability 
\begin{eqnarray}
\label{chi}
\hat\mu=1+\hat\chi,\nonumber\\
\hat\chi=\left(\matrix{\chi_{xx} &  \chi_{xy} &  0\cr
                           \chi_{yx} & \chi_{yy} & 0\cr
                           0 &  0  & \chi_{zz}\cr}\right).
\end{eqnarray}                           

Following to \cite{stockmann} 
we consider the TM modes of the Maxwell equations (\ref{maxwell}).
Substituting the permeability (\ref{chi}) into the last Maxwell equation
we have
\begin{equation}
\label{HE}
\left(\matrix{H_x\cr H_y\cr}\right)=\frac{1}{D}
\left(\matrix{\mu_{yy} & \mu_{xy}\cr \mu_{yx} & \mu_{xx}}\right)
\left(\matrix{\frac{i}{k}\frac{\partial E_z}{\partial y}\cr
\frac{-i}{k}\frac{\partial E_z}{\partial x}\cr}\right)
\end{equation}
where
$$
D=\mu_{xx}\mu_{yy}-\mu_{xy}\mu_{yx}.
$$
From this equation  one can find from the Maxwell equations (\ref{maxwell})
the wave equation (\ref{anishelm}) for electric field with eigen values equaled to
$\lambda=Dk^2$.

By the coordinate transformation
\begin{equation}
\label{trans}
\left(\matrix{x' \cr y' \cr }\right)=
\left(\matrix{-\frac{\sqrt{\mu_{xx}\mu_{yy}-(\mu_{xy}+\mu_{yx})^2/4}}
{\mu_{xx}} &0 \cr
   -\frac{\mu_{xy}+\mu_{yx}}{2\mu_{xx}} & 1 \cr} \right)
\left(\matrix{x \cr y \cr }\right)
\end{equation}
we can eliminate the cross derivatives in Eq. (\ref{anishelm}) and reduce 
the Helmholtz equation to the following on
\begin{equation}
\label{helmholz}
\nabla^2 E_z+\mu k^2 E_z=0
\end{equation}
where
\begin{equation}
\label{lambda0}
\mu=\frac{D\mu_{xx}}{\mu_{xx}\mu_{yy}-(\mu_{xy}+\mu_{yx})^2/4}.
\end{equation}
The transformation (\ref{trans}) transforms the rectangle into a particular case
of polygon as a parallelogram as shown in Fig. \ref{rombfig1} (a).

Below we consider ferrite with easy plane anisotropy, i.e.  
magnetization ${\bf M}$ in equilibrium state is perpendicular to the
anisotropy axis  ${\bf N}$. As shown in Fig. \ref{rombfig2} we direct z-axis
along the magnetization vector and x-axis along the anisotropy axis. Thereby
the $y, z$ plane is the easy plane for magnetization vector.
Assuming that resonator filed by ferrite can be considered as a thin slab we 
have the following demagnetization factors $N_x =N_y =0, N_z=1$. Moreover 
following Kittel \cite{kittel} and Lax and Button \cite{lax} we introduce an
effective demagnetizing factor as follows: $N_a=2K_a/M^2 $ which is directed along
the x-axis. Then in this specific Cartesian system of coordinates the susceptibility
has the following components

\begin{eqnarray}
\label{chixy}
\chi_{xx}(\omega)=\frac{\omega_0 \omega_M}{\omega_r^2-\omega^2},\nonumber\\
\chi_{yy}(\omega)=\frac{\omega_M(\omega_0+\omega_a-\omega_M)}{\omega_r^2-\omega^2},\nonumber\\
\chi_{xy}(\omega)=\frac{-i\omega \omega_M}{\omega_r^2-\omega^2},
\end{eqnarray}
where
\begin{eqnarray}
\label{omegas}
\omega_0=\gamma (H_0-4\pi M), \omega_M=4\pi\gamma M, \nonumber\\
\omega_r^2=\omega_0 (\omega_0+\omega_a),\nonumber\\
\omega_a=\gamma H_a, H_a=8\pi K_a/M.
\end{eqnarray}
Here $H_0$ is external constant magnetic field applied along the z-axis, i.e.
along the direction of magnetization and $H_a$ is effective anisotropy field
directed along the x-axis.

Since for this case of the susceptibility $\mu_{xy}+\mu_{yx}=0$, expression for
(\ref{lambda0}) and transformation (\ref{trans}) are simplifying as follows
\begin{equation}
\label{lambda1}
\mu=\mu_{xx}+\frac{\mu_{xy}^2}{\mu_{yy}},
\end{equation}
\begin{equation}
\label{trans1}
\left(\matrix{x' \cr y' \cr }\right)=
\left(\matrix{-\sqrt{\frac{\mu_{yy}}{\mu_{xx}}} &0 \cr
 0 & 1 \cr} \right)
\left(\matrix{x \cr y \cr }\right).
\end{equation}
If a ferrite has the anisotropy axis directed parallel to lateral sides of the 
rectangular resonator, the transformation (\ref{trans1}) maps the rectangle 
onto
rectangle. However, if the anisotropy axis of ferrite is directed differently, the
rectangle is mapping onto parallelogram as shown in Fig. \ref{rombfig1} (a). 
In particular
case of squared resonator transformation (\ref{trans1}) maps a square onto a rhombus
as shown in Fig. \ref{rombfig1} (b) with the vertex angle of the rhombus
\begin{equation}
\label{angle}
\alpha=2\arctan\sqrt{\frac{\mu_{yy}}{\mu_{xx}}}.
\end{equation}
Because of frequency dependence of the susceptibility components expressions
for eigenvalues of the wave equation (\ref{helmholz}) and the vertex angle
(\ref{angle}) are complicated. So we consider a case for which we can neglect by
frequency $\omega$ in denominator of the susceptibility (\ref{chixy}). 
Therefore we are to imply the inequality $\omega_r^2\gg\omega^2$.
In order to fulfill this inequality we can apply strong
magnetic field such as $\gamma H_0 \gg \omega$ or explore ferrites with large
anisotropy field $H_a=2K_a/M \gg \omega$. 
If to take a size of the resonator as 10cm we obtain
that, for example, the first thousand of eigen frequencies
does not exceed the frequency about $10^{11} sec^{-1}$. Therefore we can write 
inequality $H_0 \gg 10^4 Oe$. If to take
ferrite $Ni_{0.932}Co_{0.068}Fe_2O_4$ \cite{lax} we have $ 4\pi M=3475gs, 
H_a \approx 12566 Oe$.  Therefore we can neglect by frequency dependence in
denominators of the susceptibility components. The anisotropy field is
extremely large in ferrite $Ba_2Co_2Fe_{12}O_{22} (2K_a/M =2800Oe)$ , so 
upper boundary for the frequency reaches a value of order $10^{12}$.

As a result we can write the following formulas for the susceptibility
components instead of (\ref{chixy})
\begin{eqnarray}
\label{chinoomega}
\chi_{xx}\approx\frac{4\pi M}{H_0+H_a},\nonumber\\
\chi_{yy}(\omega)\approx\frac{4\pi M}{H_0-4\pi M},\nonumber\\
\chi_{xy}(\omega)\approx -\frac{i4\pi \omega M}
{\gamma(H_0+H_a)(H_0-4\pi M)}.
\end{eqnarray}
Correspondingly substituting these expressions into Eqs (\ref{lambda1}) and
(\ref{angle}) we have
\begin{eqnarray}
\label{mu0}
\mu\approx 1+\frac{4\pi M}{H_0+H_a}-\frac{16\pi^2 M^2\omega^2}
{\gamma^4(H_0+H_a)^2(H_0-4\pi M)^2},\\
\label{angle0}
\alpha\approx 2\arctan\sqrt{\frac{H_0}
{(H_0-4\pi M)(1+\frac{4\pi M}{H_0+H_a})}}.
\end{eqnarray}
One can see that if the external magnetic field $H_0$ and the anisotropy field
$H_a$ are both strong we obtain from (\ref{mu0}) and (\ref{angle}) 
$\mu\approx 1, \alpha\approx \pi/2$. It means that the strong magnetic field 
applied along the magnetization of ferrite (z-axis) diminishes effect of 
anisotropy. Thereby we assume that only the anisotropy field is strong to obtain
\begin{eqnarray}
\label{mu1}
\mu\approx 1,\quad \mu_{xx}\approx 1\\
\label{angle1}
\alpha\approx 2\arctan\sqrt{\frac{H_0}{H_0-4\pi M}}.
\end{eqnarray}
On the one hand, formula (\ref{mu1}) shows that eigen frequencies of 
filed resonator coincide with those of empty resonator. On the other
hand  formula (\ref{angle1}) gives a remarkable possibility to change the vertex
angle of the rhombus by external magnetic field.

While integrable, for example, rectangular billiard, and chaotic,
for example, the Sinai billiard, represent two extreme billiards,
the rhombus with at least one angle in the form $m\pi/n (m\neq 1)$
belongs to pseudo integrable systems 
\cite{eckhard,biswas,shudo,gremaud,bogomolny}.
Integrable billiard generally leads to uncorrelated energy levels (Poisson
statistics) and chaotical billiard corresponds to the Wigner-Dyson statistics
\cite{stockmann}. Rhombus billiards are pecular as they are pseudo integrable
systems and for this reason their statistical properties belong to another
class of universality \cite{parab}.
Using the boundary element method \cite{riddel} Shudo and Shimizu \cite{shudo}
have found that the data of nearest-neighbour level-spacing
distributions are described by the Brody distribution
\begin{eqnarray}
P_{\beta}(s)=As^{\beta}\exp(-a s^{1+\beta}),\nonumber\\
\label{brody}
A=(1+\beta)a,~
a=\left[\Gamma\left\{\frac{2+\beta}{1+\beta}\right\}\right].
\end{eqnarray}
The Brody distribution is is a semiempirical interpolation between
Poisson and Wigner-Dyson distributions \cite{brody}.
Shudo and Shimizu compared the Brody parameter $\beta$ for
irrational angles with that for the rational angles and found that
the differences observed are very small. However, the spectral
rigidity reveals the possibility of the rationality of vertex angle
being an important signature of the level-spacing distribution.
Bogomolny {\it et al} \cite{bogomolny} performed extensive numerical calculations
for $\alpha=\pi/n, n=5, 7, ..., 30$ and typically 20000 energy levels and found
that the nearest-neighbour level-spacing distribution is described by
the semi-Poisson statistics
\begin{equation}
P_{\beta}(s)=4se^{-2s}.
\label{semi}
\end{equation}
Biswas and Jain \cite{biswas} numerically considered the
amplitude distributions $P(\psi)$ of
eigenfunctions of the $\pi3$ rhombus. They have found that  eigen
functions even-even and even-odd relative to the $x, y$ axes 
(Fig. \ref{rombfig2}) 
for this rhombus $P(\psi)$ have the Gaussian distribution in
contrast to the odd-odd and odd-even eigenfunctions 
which are identical to those of
an equilateral triangle. These eigenfunctions are
\cite{brack}
\begin{eqnarray}
\label{triangle}
\psi_{\bigtriangleup}=(\cos\sin)\left[\frac{2\pi (2m-n)x}{3L}\right] 
\sin\left[\frac{2\pi ny}{\sqrt{3L}}\right]\nonumber\\
 -(\cos\sin)\left[\frac{2\pi (2n-m)x}{3L}\right]
 \sin\left[\frac{2\pi my}{\sqrt{3L}}\right]
 (\cos\sin)\left[\frac{-2\pi (m+n)x}{3L}\right]
 \sin\left[\frac{2\pi (m-n)y}{\sqrt{3L}}  \right]
\end{eqnarray}
with the eigen energies $E_{mn}=E_{\Delta}(n ^2+m^2-mn)$
where $E_{\Delta}=16\hbar^2\pi^2/(18mL^2)$ with $m\geq 2n,\,
n=1,2,3,\ldots$ and $L$ is the length of triangle side. This size related to
The length of square side $l$ related to this length by relation 
$L=l/\sqrt{2}\sin(\alpha/2)$ (Fig. \ref{rombfig1}).
From equations $\alpha=2\pi/3$ and (\ref{angle1}) we obtain that this
integrable case of the anisotropic square resonator takes place if
\begin{equation}
\label{pi3}
H_0=2\pi M
\end{equation}
otherwise the resonator is non integrable one.

This work has been supported by  Russian Foundation for Basic Research 
Grant 01-02-16077 and the Royal Swedish Academy of Sciences (A.F.S.).

\begin{figure}
\caption{Mapping of rectanglar billiard described by the anisotropic Helmholtz
equation(\ref{anishelm}) on to polygon (a) (transformation (\ref{trans}))
and of square billiard on to a rhombus (b) (transformation (\ref{trans1})).}
\label{rombfig1}
\end{figure}

\begin{figure}
\caption{Schematical view of the rectangular resonator filed with
ferrite where ${\bf M}$ is the magnetization of ferrite and
${\bf N}$ is the anisotropy field.}
\label{rombfig2}
\end{figure}


\begin{thebibliography}{99}
\bibitem{mcdonald}
S.W. McDonald and A.N. Kaufman, Phys. Rev. Lett. {\bf 42}, 1189 (1979).
\bibitem{bohigas}
O. Bohigas, M.-J. Giannoni, and C. Schmit, Phys. Rev. Lett. {\bf 52}, 279 (1984).
\bibitem{mcdonald1}
S.W. McDonald and A.N. Kaufman, Phys. Rev. A{\bf 37}, 3067 (1988).
\bibitem{stockmann}
H.-J. St\"okmann, Quantum Chaos: An Introduction (Cambridge
University Press, Cambridge, UK, 1999).
\bibitem{stockmann1}
H.-J. St\"okmann and  J. Stein, Phys. Rev. Lett. {\bf 64}, 2215 (1990); {\it ibid}
{\bf 68}, 2867 (1992).
\bibitem{sridhar}
S. Sridhar, Phys. Rev. Lett. {\bf 67}, 785 (1991).
\bibitem{sridhar1}
S. Sridhar, E.J. Heller,  Phys. Rev. A{\bf 46}, R1728 (1992).
\bibitem{alt}
H. Alt, H.-D. Gr\"af, H.L. Harney, R. Hofferbert, H. Lengeler, A. Richter, 
P. Schardt, and H.A. Weidenm/"uller, Phys. Rev. Lett. {\bf 74}, 62 (1995).
\bibitem{alt1}
H. Alt, C. Dembovski, H.-D. Gr\"af, R. Hofferbert, H. Rehfeld, A. Richter
and C. Schmidt, Phys. Rev. E{\bf 60}, 2851 (1999).
\bibitem{so}
P. So, S.M. Anlage, E. Ott, and R.N. Oerter, Phys. Rev. Lett. {\bf 74}, 2662 (1995).
\bibitem{stoffregen}
U. Stoffregen, J. Stein, H.-J. St\"okmann, M.Ku\'s, and F.Haake,
Phys. Rev. Lett. {\bf 74}, 2666 (1995).
\bibitem{kittel}
C. Kittel, Phys. Rev. {\bf 73}, 155 (1948).
\bibitem{lax}
B. Lax, K.J. Button, Microwave Ferrites and Ferrimagnetics, N.Y., 1962.
\bibitem{eckhard}
B.Eckhard, J.Ford and F.Vivaldi, Physica {\bf 13D}, 339 (1984).
\bibitem{biswas}
D. Biswas and S.R. Jain, Phys. Rev. A{\bf 42}, 3170 (1990).
\bibitem{shudo}
A. Shudo and Y. Shimizu, Phys. Rev. E{\bf 47}, 54 (1993).
\bibitem{gremaud}
B.Gr\'emaud  and S.R.Jain, J. Phys. A: Math. Gen. {\bf 31}, L637 (1998).
\bibitem{bogomolny}
E.B. Bogomolny, U.Gerland, and C.Schmit, Phys. Rev. E{\bf 59}, R1315 (1999).
\bibitem{parab}
H.D. Parab and S.R.Jain, J. Phys. A: Math. Gen. {\bf 29}, 3903 (1996).  
\bibitem{riddel}
R.J. Riddel, Jr., J. Comp. Phys. {\bf 31}, 21 (1979); {\bf 31}, 42
(1979).
\bibitem{brody}
T.A. Brody, Lett. Nuovo Cimento, {\bf 7}, 482 (1973).
\bibitem{brack} 
M. Brack and R.K. Badhuri, "Semiclassical Physics", Addison-Wesley, (1997)
\end{thebibliography}
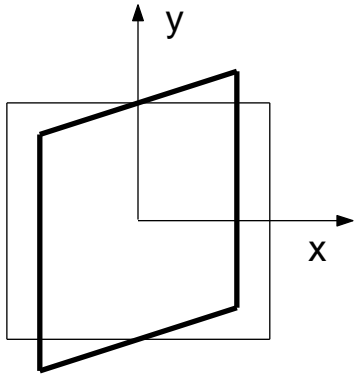
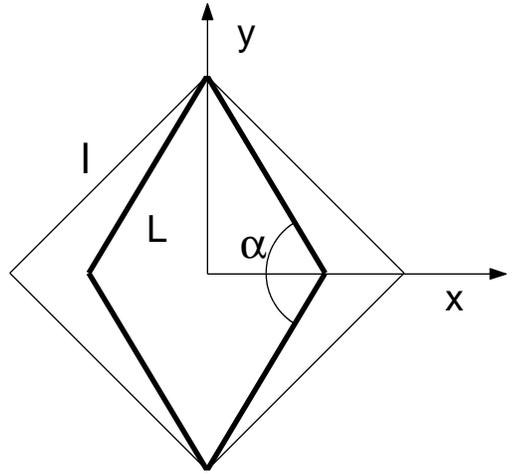
\end{document}